\documentclass[12pt,dvips]{article}
\usepackage{color}
\usepackage{bm}
\usepackage{url}
\usepackage{hyperref}
\usepackage{graphicx}
\usepackage{amsmath}
\usepackage{amssymb}
\usepackage{makeidx}
\title{Polarized light ions and spectator nucleon 
tagging at EIC\footnote{Proceedings of DIS 2014, 
XXII.~International Workshop on Deep-Inelastic 
Scattering and Related Subjects, University of Warsaw, Poland,
April 28 - May 2, 2014}
\vspace{-2ex}}
\author{V.~Guzey$^1$, D.~Higinbotham$^2$, Ch.~Hyde$^3$, 
P.~Nadel-Turonski$^2$, \\[-.5ex]
K.~Park$^3$, M.~Sargsian$^4$, 
M.~Strikman$^5$, C.~Weiss$^2$ \\[1ex]
\small $^1$\ Petersburg Nuclear Physics Institute, Gatchina, 188300, 
Russia \\[-.7ex]
\small $^2$\ Jefferson Lab, Newport News, VA 23606, USA \\[-.7ex]
\small $^3$\ Old Dominion University, Norfolk, VA 23529, USA \\[-.7ex]
\small $^4$\ Florida International University, Miami, FL 33199, USA \\[-.7ex]
\small $^5$\ Pennsylvania State University, University Park, PA 16802, USA
\vspace{-2ex} }
\begin{document}
\maketitle
\begin{abstract}
An Electron--Ion Collider (EIC) with suitable forward detection 
capabilities would enable a unique experimental program of deep--inelastic 
scattering (DIS) from polarized light nuclei (deuterium $^2$H, helium $^3$He) 
with spectator nucleon tagging. Such measurements promise significant advances 
in several key areas of nuclear physics and QCD: (a)~neutron spin structure, 
by using polarized deuterium and eliminating nuclear effects through on-shell
extrapolation in the spectator proton momentum; 
(b)~quark/gluon structure of the bound nucleon at $x > 0.1$
and the dynamical mechanisms acting on it, by measuring the spectator 
momentum dependence of nuclear structure functions; 
(c)~coherent effects in QCD, by exploring 
shadowing in tagged DIS on deuterium at $x \ll 0.1$. 
The JLab MEIC
design (CM energy $\sqrt{s} = 15-50$ GeV/nucleon, luminosity
$\sim 10^{34} \, \textrm{cm}^{-2} \, \textrm{s}^{-1}$) provides 
polarized deuterium beams and excellent coverage and resolution for
forward spectator tagging. We summarize the physics topics, 
the detector and beam requirements for spectator tagging, 
and on-going R\&D efforts.
\end{abstract}
\newpage
The Electron--Ion Collider (EIC) proposed as a 
next-generation facility for nuclear physics would offer unprecedented
capabilities for high--energy scattering on light nuclei (deuterium ${}^2$H, 
helium ${}^3$He, \ldots), including polarized beams. The medium--energy 
EIC designs presently developed feature center--of--mass energies in the range 
$\sqrt{s} \sim$ 15-50 GeV/nucleon at luminosities up to $\sim 10^{34} \, 
\textrm{cm}^{-2} \, \textrm{s}^{-1}$ \cite{EIC-designs}. 
Measurements of deep--inelastic 
scattering (DIS) and related processes on light nuclei in this kinematic 
region address several basic questions of nuclear 
physics:\footnote{For a general overview of the medium--energy EIC physics 
program, including proton beams, see e.g.\ Ref.~\cite{Accardi:2011mz}.}
\begin{itemize}
\item[A)] \textit{Neutron spin structure.} What are the spin structure 
functions of the neutron and their $Q^2$ dependence? This information is 
essential for the flavor decomposition of the quark spin densities
in the nucleon. It also impacts on the determination of the gluon spin 
density, as proton data alone are not sufficient to separate quarks and 
gluons. The isovector structure function $g_{1p} - g_{1n}$
exhibits especially simple QCD evolution (insensitive to gluons),
permits accurate separation of leading and higher--twist contributions,
and is needed to test the Bjorken sum rule.
\item[B)] \textit{Bound nucleon structure.} What are the quark/gluon
distributions of the bound nucleon at $x > 0.1$? By what mechanisms are they 
modified in the nuclear medium? How does the modification depend on the
nuclear configuration (off--shellness, strength of interaction)? 
Answering these questions will help
to understand the QCD origin of the short--range $NN$ interaction
and the role of non-nucleonic degrees of freedom in nuclei.
\item[C)] \textit{Coherent scattering in QCD.} How does the quantum effect
of coherence in high-energy scattering manifest itself in QCD? How does it
influence the gluon and quark densities seen by a short--distance probe
at $x \ll 0.1$? Coherence results in shadowing, a basic prediction of 
QCD which can be observed experimentally. It also determines how rapidly
the regime of high gluon densities (saturation) is approached at small $x$ .
\end{itemize}
In trying to answer these questions on the basis of actual nuclear DIS
data one faces considerable challenges. In the extraction of free neutron 
structure one must eliminate the effects of nuclear binding and final--state 
interactions, and accurately account for the neutron polarization in 
the nucleus. In the study of bound nucleon structure one needs to control 
the nuclear environment, which is compounded from different types of 
configurations that one would like to separate (mean field, short-range 
correlations). In the search for coherent effects one wants to identify 
unambiguous signatures of coherence and, ideally, separate contributions
involving $N = 2, 3, \ldots$ nucleons. While all this can be accomplished 
partly with theoretical calculations, it is clear that more experimental 
control is needed in the analysis of nuclear DIS data. Two new 
experimental tools that become available with the EIC would greatly 
help one to address the challenges.

One essential tool are deuterium beams, especially polarized deuterium, 
as would be available for the first time with JLab MEIC, thanks to the 
figure--8 shape of the ion ring designed to compensate the effect of spin 
precession. The deuteron is the simplest nucleus ($A = 2$); its wave 
function is known well up to large relative momenta $\sim$ few 100 MeV, 
including the light-front wave function describing microscopic nuclear 
structure as probed in high-energy scattering 
processes \cite{Frankfurt:1981mk}. 
The deuteron has spin 1 and is mostly in the $L = 0$ configuration 
(S--wave), with a small admixture of $L = 2$ (D--wave), such that the 
proton and neutron are spin--polarized and their degree of polarization 
is known very well. Because there are only two nucleons the possibilities 
for final--state interactions are limited; in configurations where they
can happen they can be estimated using theoretical models. Finally, 
at small $x$ the deuteron allows one to study coherent effects exactly in the 
$N = 2$ system, greatly simplifying the theoretical analysis.

The other tool is the detection of spectator nucleons emerging from the 
high--energy scattering process (``spectator tagging''). In collider 
experiments the 
spectator nucleons carry fraction $\sim 1/A$ of the ion beam momentum and can 
be detected with appropriate forward detectors (see below). The technique 
is uniquely suited to colliders: there is no target material absorbing 
low-momentum nucleons, and it can be used with polarized ion beams
(longitudinal and transverse).
Spectator tagging is especially powerful in scattering on 
deuterium. It allows one to positively identify the active nucleon
(e.g., DIS on the $n$ with a $p$ spectator detected) and control its
quantum state through measurement of the recoil momentum.
Spectator tagging with unpolarized deuterium was explored in
a pioneering fixed-target experiment at JLab with 6 GeV beam energy
(CLAS BoNuS detector, covers recoil momenta $p_R \gtrsim 70\, \textrm{MeV}$) 
\cite{Tkachenko:2014byy} and will be studied further at 11 GeV.

Spectator tagging with polarized deuterium at EIC represents a unique 
combination that would qualitatively advance our understanding of 
nuclear effects in DIS and answer the fundamental nuclear physics questions 
listed above. It enables a theoretical analysis of nuclear DIS at a level 
of precision that is commensurate with the expected quality of the data. 
In this note we briefly summarize the physics impact, the detector and beam 
requirements, and on-going R\&D efforts.

\begin{figure}
%\centering
\parbox[c]{0.38\textwidth}{
\begin{center}
\includegraphics[width=0.26\textwidth]{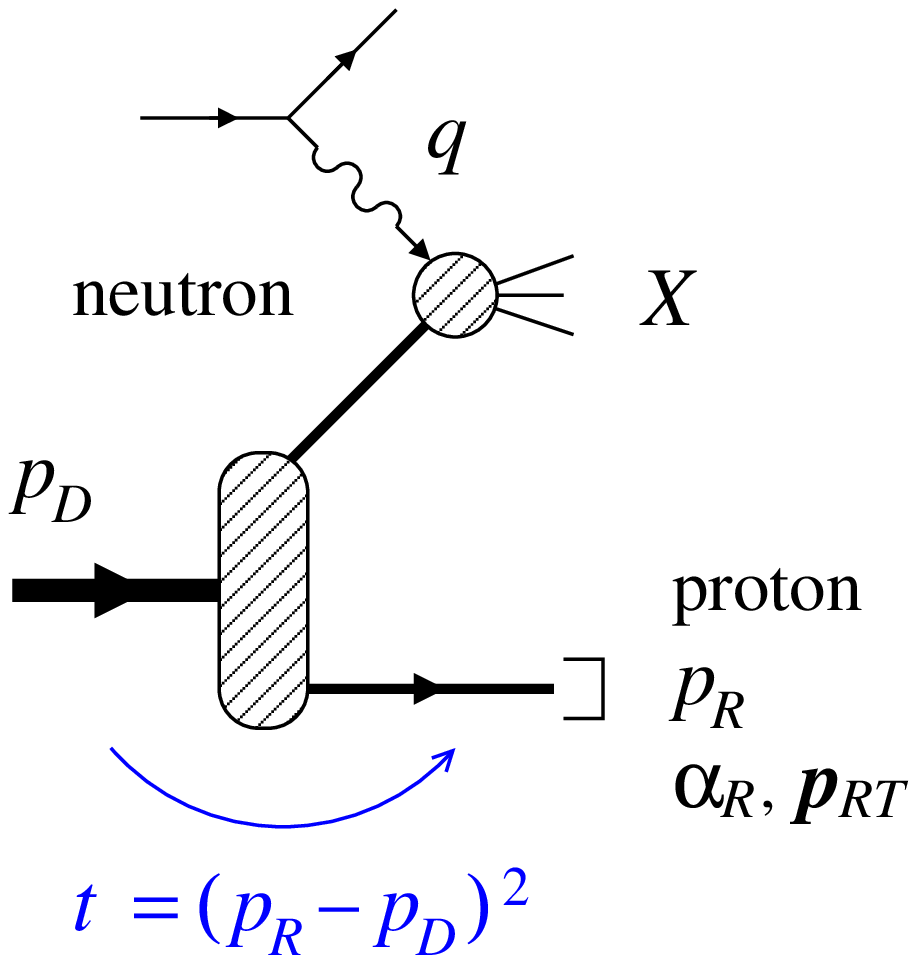}
\end{center}
\vspace{-6ex}
{\small (a)}
\vspace{2ex}
\caption{\small (a) Tagged DIS $e+ D \rightarrow e'+p+X$. The scattering
amplitude has a pole at $t = M_N^2$, given by the impulse approximation
diagram. (b) On-shell extrapolation $t \rightarrow M_N^2$ 
in a simulated measurement with EIC \cite{LDRD}. 
\label{fig:neutron}
}}
\hspace{0.04\textwidth}
\parbox[c]{0.57\textwidth}{
\includegraphics[width=0.57\textwidth]{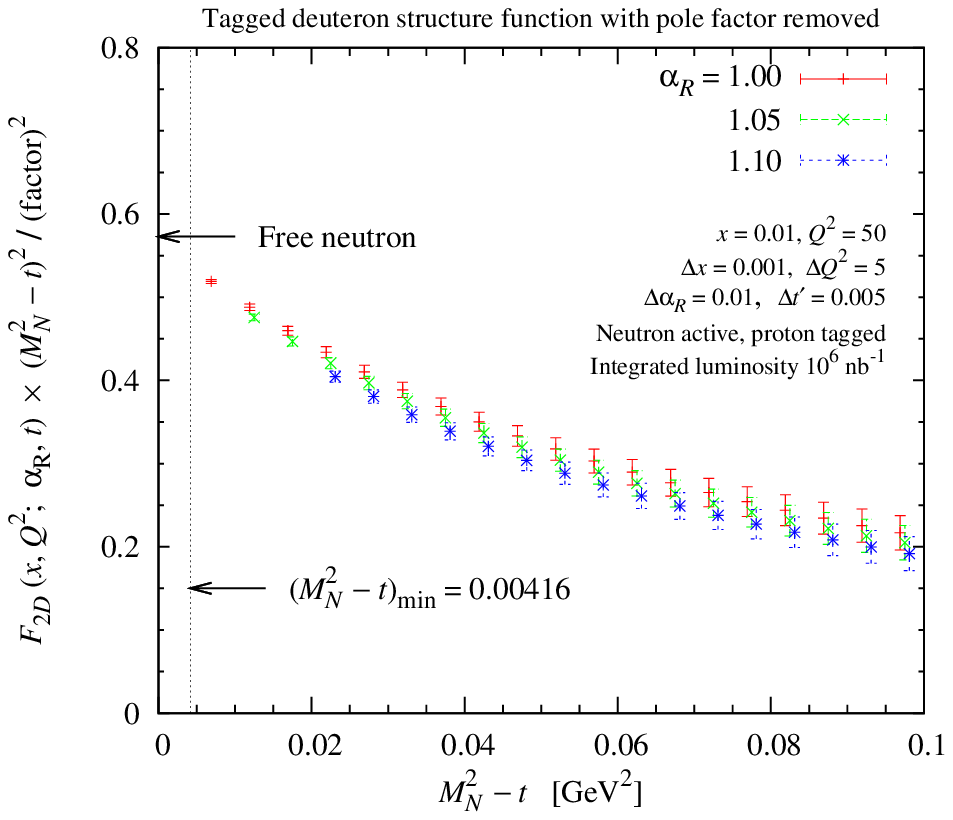}
\\[-5ex]
{\small (b)}}
\end{figure}

{\bf Neutron structure.} Spectator tagging with deuterium provides a 
model-independent method for determining the DIS structure functions of 
the free neutron. One measures the conditional DIS cross section 
$e + D \rightarrow e' + p + X$ (here $D \equiv$ $^2$H) 
as a function of the recoil proton momentum,
described by the light--cone fraction $\alpha_R \equiv 
2 (E_R + p_R^z)/(E_D + p_D^z)$ and the transverse momentum $\bm{p}_{RT}$
(the components refer to a frame in which the deuteron and virtual 
photon momenta are collinear and define the $z$--direction).
Another important variable is the invariant 4--momentum transfer between the
deuteron and the recoil proton, $t \equiv (p_R - p_D)^2$. As a function of
$t$ the scattering amplitude has a pole at $t = M_N^2$ (unphysical 
region) corresponding to nucleon exchange in the $t$--channel
(see Fig.~\ref{fig:neutron}a). The residue at
the pole is, up to a constant factor representing deuteron structure,
given by the structure function of the free neutron, evaluated at the
argument $\tilde x = x/(2 - \alpha_R)$. Nuclear binding and final--state 
interactions only affect the amplitude away from the pole, but not
the residue at the pole \cite{Sargsian:2005rm}.
%
% FIGURE
%

To extract the free neutron structure function one measures the tagged 
cross section over a range of $t$, removes the pole factor 
$1/(t - M_N^2)^2$, and extrapolates to 
$t \rightarrow M_N^2$ \cite{Sargsian:2005rm}. 
The pole in $t$ is extremely close to the physical region 
(the distance is proportional to the deuteron binding energy, 
$\epsilon_D M_D$) so that the extrapolation can be performed with 
great accuracy. The method is analogous to the Chew--Low extrapolation 
used to extract pion structure from $\pi N$ scattering data. 
Figure~\ref{fig:neutron}b shows a simulated on-shell extrapolation with 
MEIC pseudodata ($s_{eD} = 1000\, \textrm{GeV}^2$, integrated luminosity 
$10^{6}\, \textrm{nb}^{-1}$) \cite{LDRD}. One sees that the extrapolation is 
very smooth. Comparison of the data at different recoil light-cone
fraction $\alpha_R$ allows one to test universality of the nucleon pole. 
Critical to the success of the method is the ability to measure the recoil 
momentum with complete coverage down to $p_{RT} = 0$ and resolution 
$\Delta p_{\rm RT} \lesssim \textrm{20}\, \textrm{MeV}$
and $\Delta \alpha_R \lesssim 10^{-3}$ (see below).

Spectator tagging with on-shell extrapolation can be used for precision 
measurements of $F_{2n}$ (and, indirectly, the difference $F_{2p} - F_{2n}$), 
which determine the flavor structure of the nucleon sea and validate models
of nuclear effects. Even more importantly, it can be used to extract 
the neutron spin structure function $g_{1n}$ from measurements with polarized 
deuterium beams. It represents the cleanest and most model-independent 
method to determine the neutron spin structure. The impact
of neutron data on the flavor decomposition of the quark spin and the
determination of the gluon polarization has been discussed
in the literature. The method can also be extended
to semi-inclusive DIS on the neutron (flavor decomposition of transverse
momentum distributions and spin--orbit effects) and exclusive final states
(generalized parton distributions).

{\bf Bound nucleon structure.} Spectator tagging with polarized deuterium
also offers a unique method for studying the modification of the nucleon's
quark/gluon structure in the nucleus. Measurements of ratios of inclusive 
nuclear structure functions at $x > 0.25$ show a distinctive nuclear 
dependence (``EMC effect''), whose origin has been the subject of much 
theoretical speculation. Spectator tagging 
would allow one to study the nuclear modification as a function 
of the recoil nucleon momentum, extending the measurements of 
Fig.~\ref{fig:neutron}b over a broad range of momentum transfers 
up to $t - M_N^2 \sim -0.5\, \textrm{GeV}^2$. 
In this way one could directly reveal the connection between the 
nuclear modification and short--range $NN$ 
correlations, resolving a basic question in the interpretation.
By combining spectator proton tagging in DIS on deuterium with 
conventional flavor tagging (semi-inclusive DIS) one could 
in addition determine whether the (polarized) $u$ and $d$ quarks in the 
neutron are modified in the same way. Finally, because 
of the wide kinematic coverage in $Q^2$ one could map out the $Q^2$ evolution
of the nuclear structure functions at $x > 0.1$ and separate the modification
of quark and gluon distributions. Altogether, these measurements would
enable a new level of understanding of the modification of the
nucleon's partonic structure in nuclei. This program relies essentially
on the high luminosity of the EIC, as the tagged deuteron cross sections 
decrease very fast when increasing the spectator momentum.

%
% FIGURE
%
\begin{figure}
\parbox[c]{0.4\textwidth}{
\hspace{.04\textwidth}
\includegraphics[width=0.32\textwidth]{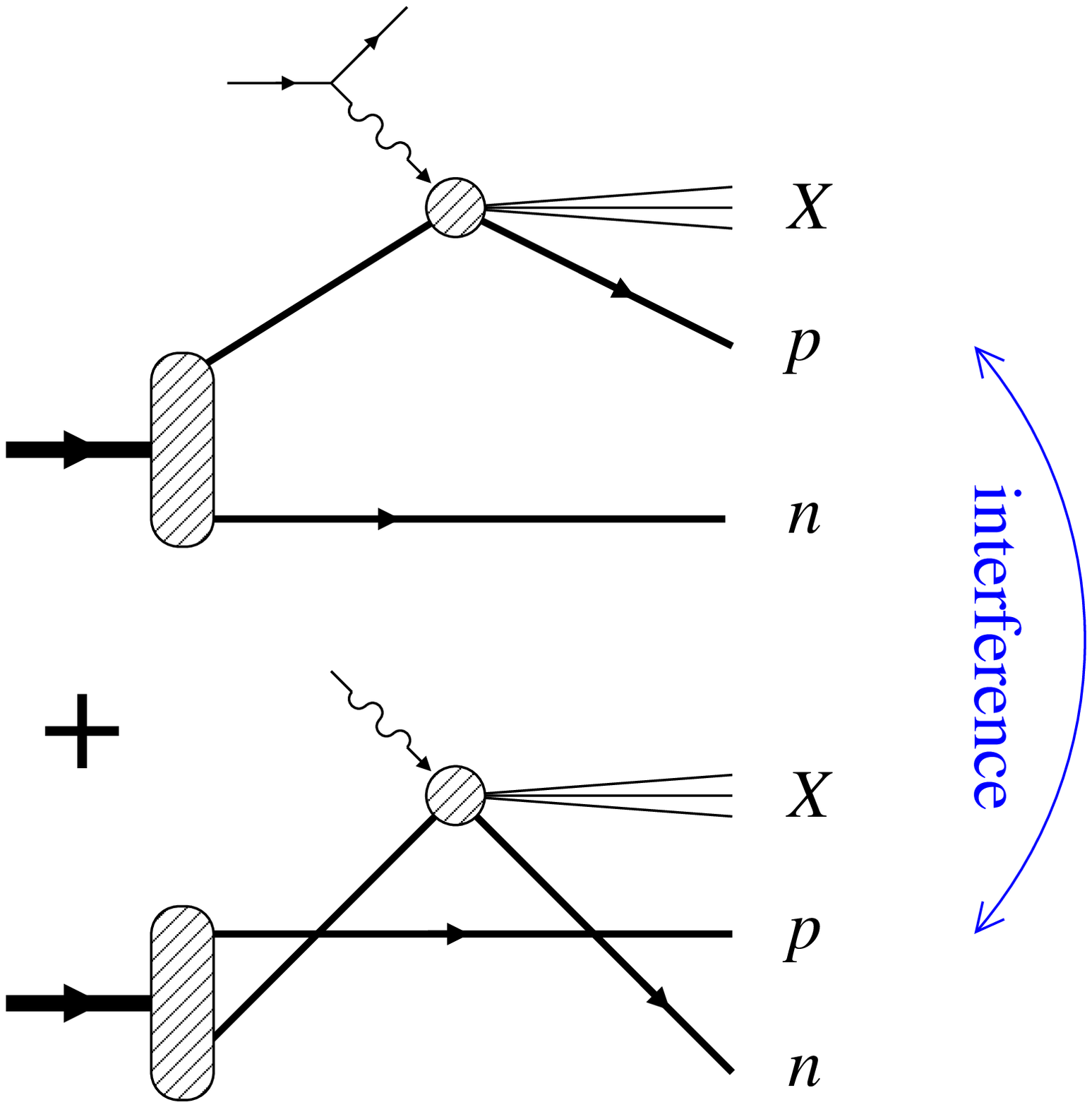}
\\[-2ex]
{\small (a)}
}
\hspace{0.1\textwidth}
\parbox[c]{0.47\textwidth}{
\includegraphics[width=0.47\textwidth]{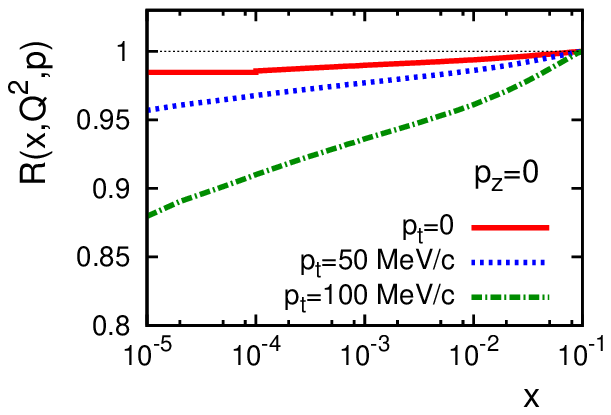}
\\[-2ex]
{\small (b)}
}
\caption{(a) Shadowing in tagged DIS on deuterium. Diffractive scattering
on the $p$ and $n$ causes interference between the two amplitudes. 
(b) Theoretical prediction for the shadowing ratio 
$R = F_{2D}/(F_{2p} + F_{2n})$ as a function of $x$ and the recoil proton
transverse momentum \cite{Frankfurt:2011cs}. 
\label{fig:coherence}}
\end{figure}
{\bf Coherent scattering in QCD.} In DIS at $x \ll 0.1$ the longitudinal
extent of the interaction becomes much larger than the typical internucleon 
distances in nuclei, so that the high-energy probe can interact coherently 
with all the nucleons lined up along its path, giving rise to distinctive 
phenomena such as shadowing. Attempts to observe these phenomena 
experimentally have 
so far relied mostly on inclusive DIS from heavy nuclei. In DIS on light 
nuclei with spectator tagging coherence manifests itself in novel ways 
and can be studied with much better experimental control.

At $x \ll 0.1$ there is a significant probability for DIS on the nucleon
to produce a diffractive final state, where the nucleon remains intact and
recoils with a momentum transfer $\sim$ few~100~MeV (see 
Ref.~\cite{Frankfurt:2005mc} for a discussion of the HERA results).
In DIS on deuterium such diffractive scattering can happen on the 
proton or the neutron, causing quantum-mechanical interference between the 
two amplitudes with the same final state (see Fig.~\ref{fig:coherence}a). 
In inclusive DIS, $e + D \rightarrow e' + X$,
the interference gives rise to leading--twist nuclear shadowing and
can be calculated as an integral over the deuteron wave function
\cite{Frankfurt:2011cs}. In tagged DIS, $e + D \rightarrow e' + p + X$
(or double--tagged DIS, $e + D \rightarrow e' + p + n + X$), the interference
effect can be observed directly as a function of the recoil momentum
(see Fig.~\ref{fig:coherence}b), especially at larger transverse momenta 
$p_{RT} \sim \, 100\, \textrm{MeV}$, where it can be as large as
10-20\%. Such measurements would enable detailed tests of the theoretical
description of coherence and shadowing at small $x$. Besides its intrinsic 
interest this would benefit the extraction of parton densities from 
nuclear DIS data and the phenomenology of the black--disk regime 
(unitarity limit) at small $x$, as shadowing influences quantitatively 
how rapidly this regime is approached in nuclei. The diffractive 
parton densities needed as input to the shadowing calculations can
be measured in $ep$ scattering with EIC in the same kinematics.

An attractive feature of the deuteron is that one can study coherence
strictly in the $N = 2$ system and does not have to deal with multiple 
scattering from $N > 2$ nucleons. By going from $^2$H to $^3$He 
one can then switch on the $N = 3$ term in the multiple scattering series.
This stepwise approach represents an interesting complement to coherence
studies with heavier nuclei.\footnote{Another observable potentially 
sensitive to coherence is the deuteron's tensor--polarized structure 
function $b_1$, which is absent in scattering from a single free 
nucleon \cite{Frankfurt:1981mk}.}

{\bf Detector and beam requirements.} Spectator tagging
with EIC requires integrated forward detectors with (a) complete coverage 
for protons with low recoil momenta relative to beam momentum per nucleon:
$p_{RT} < 200 \, \textrm{MeV}, 
\; p_{R\parallel}/p(\textrm{beam}) \sim 0.8-1.2$;
(b) sufficient recoil momentum resolution: 
$\Delta p_{RT} \lesssim 20 \, \textrm{MeV}, \;
\Delta p_L/p_L \sim 10^{-4}$; (c) ideally, also neutron detection with 
reasonable angular and position resolution. The MEIC interaction region 
and forward 
detection system have been designed specifically for this purpose and 
provide fully sufficient capabilities for the physics program outlined
here \cite{meic_forward}. Spectator tagging also requires that the
intrinsic momentum spread in the ion beam be sufficiently small to
allow for accurate reconstruction of the actual recoil momentum at
the interaction vertex. Simulations show that with the MEIC beam 
parameters the ``smearing'' of the kinematic variables is very moderate 
and does not substantially affect the physics analysis (in 
$t$ the uncertainty is of the order $\sim 0.005\, \textrm{GeV}^2$ --- 
the bin size in Fig.~\ref{fig:neutron}b) \cite{LDRD}.

An R\&D program in under way at JLab to develop simulation tools
for spectator tagging with EIC (cross section models, event generators)
and demonstrate the feasibility of such measurements~\cite{LDRD}. 
The tools are being made available to users and can applied to 
a variety of processes of interest. Information about available
resources may be obtained from the authors.

Notice: Authored by Jefferson Science Associates, 
LLC under U.S.\ DOE Contract No.~DE-AC05-06OR23177. The U.S.\ Government 
retains a non--exclusive, paid--up, irrevocable, world--wide license to 
publish or reproduce this manuscript for U.S.\ Government purposes.
\end{document}